# *What was in the apparatus before the click of the detector?*

Sofia Wechsler

## Abstract
When a quantum system is described by a superposition of wave-packets, each wave-packet traveling on a separate path, a commonly asked question is *why* only one of the wave-packets is able to trigger a click in a detector. In the second half of the last century many scientists considered the possibility that not all these wave-packets are identical. Namely, that there exist "full waves" and "empty waves". The two types of waves were supposed to be identical in the sense that they are able to produce interference when crossing one another, however, the full wave was supposed to be able to trigger a click in a detector, while the empty wave was supposed to leave the detector silent. The present text describes an experiment which, for explaining the results, seems to necessitate the hypothesis of full and empty waves.

## Abbreviations:
2-wave = two-particle wave
dBB = de Broglie-Bohm
DC = down-conversion
QM = quantum mechanics
s-i = signal-idler
UV = ultraviolet
w-f = wave-function
w-p = wave-packet

## 1. Introduction

Although the quantum mechanics (QM) succeeded to explain a wide range of phenomena of the microscopic world, for almost a century the most basic features of a quantum systems are still not clarified. How does look like this system? What is the wave-function (w-f), some real wave traveling in our apparatus, or just a mathematical tool for predicting probabilities? In the latter case, how looks like the "creature" that travels in our apparatus?

A very puzzling property of the quantum systems is the quantum superposition. When the w-f consists in a couple of wave-packets traveling in different regions of the space, "which-way" experiments show that only one of the wave-packets (w-ps) is able to trigger a detector. It seems therefore that only one of the w-ps exists in reality. But this is a false impression. If instead of placing detectors on the paths of the w-ps, the w-ps are deflected and brought to intersect one another, in the intersection region appear interference fringes. This is evidence that both the intersecting w-ps exist in reality. *Then, why only one of them impresses a detector?*

An appealing answer would be that the w-ps differ in their possibility to cause a detection: one of the w-ps possesses this possibility – and is usually called in the literature "full wave" – while all the other w-ps are "empty waves", i.e. do not have this possibility. A wide debate unfolded around these matters in the end of the last century, [1 – 15]. Historically, the terms "empty wave" and "full wave" came from the idea that in the microscopic world there exists a substructure element, not appearing in the QM formalism, an entity floating inside the w-f. Outside the volume occupied by that entity, the w-f was supposed to be "empty" i.e. of no effect on the detector.[1]
The best elaborated expression of this idea was the de Broglie-Bohm (dBB) interpretation of the QM, [1 – 3]. This interpretation is based on the assumption that there exists a "particle" inside the w-f, and the particle travels along a continuous trajectory together with the w-f and inside it. If the w-f consists of several w-ps, the particle travels with one of them, and the rest of the w-ps are considered some sort of really existing waves, though to which the detectors are insensitive. However, a strong argument was brought against the dBB mechanics in [18].[2]

What was challenged in [18] wasn't the existence of a particle as a substructure of the QM formalism. The possibility of a continuous trajectory for the alleged particle was proved incompatible with the QM predictions. The proof can be immediately generalized for ruling out continuous trajectories for full waves. Still, in [18] the door is left open for assuming the existence of a particle that jumps between the w-ps. Such an interpretation of the QM was proposed by S. Gao, namely that there exists a particle in random, discontinuous motion (RDM) [19 – 21]. Regrettably, the random motion is incompatible with the correlations in entanglements. A problem appears if one takes into account that experimenters have *free will* and can choose, independently of one another, at which time to measure the particles and which type of test to perform. An analysis of the RDM interpretation, of its advantages and weaknesses, was done in [22].

The rest of the article has the following line: in the second section an experiment is described, and the results are examined in section 3. The section 4 analyses two particular cases of the experiment in the light of the idea of full and empty waves, and finds that this idea offers a very plausible explanation. The section 5 contains conclusions.

## 2. An experiment with down-conversion pairs of photons

A beam of UV photons emitted by a laser, is split by a 50%-50% beam-splitter BS, figure 1. There result a transmitted beam, $|1_p\rangle$, and a reflected beam $|1_{p'}\rangle$. These beams land on two identical non-linear crystals, X and **X**' respectively. In each crystal, a tiny fraction, $|\alpha|^2$, from the incident UV beam, undergoes down-conversion (DC) to signal-idler (s-i) pairs,

$$|1_p,0,0\rangle \rightarrow \beta\,|1_c,0,0\rangle + \alpha\,|0,1_s,1_i\rangle\,, \tag{1}$$

$$|1_{p'},0,0\rangle \rightarrow \beta\,|1_{c'},0,0\rangle + \alpha\,|0,1_{s'},1_{i'}\rangle\,, \tag{2}$$

---

[1] The idea of two types of waves may remind to the reader the transactional interpretation of the QM, due to J.G. Cramer [16]. [17], in which two waves were assumed to determine the measurement result. One wave was assumed to propagate from the source to the detectors forward in time, and the other wave was assumed to propagate backward in time from the detectors to the source. In the hypothesis of full and empty waves, nothing propagates backward in time.

[2] The general line of the proof in [18] is that from a source of particles there are three paths on which the particle may go. It is proved that the particle should have taken at once two of these paths. However, if a particle follows a continuous trajectory between source and detector, it may travel only on one path. Bohm's interpretation of the QM is based on the assumption of continuous trajectories.

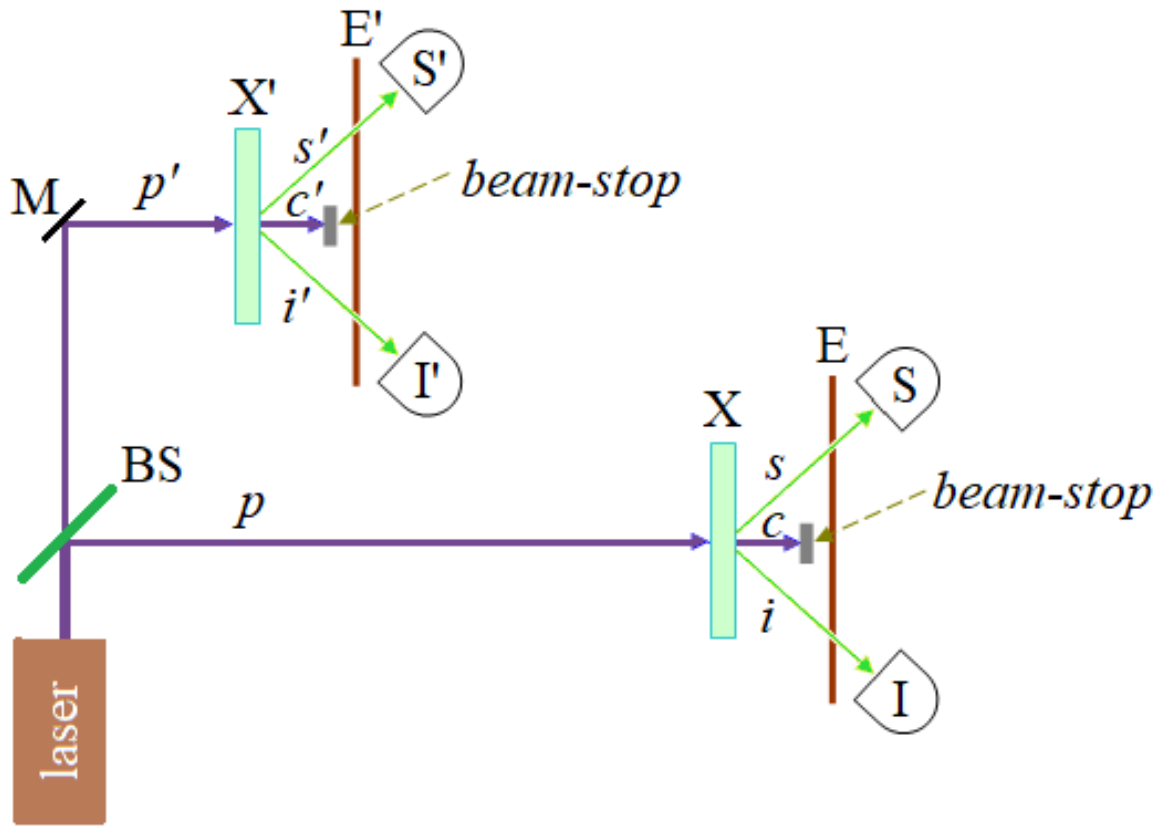

Figure 1. DC-pair production in two non-linear crystals.
See explanations in the text.

with $|\beta|^2 + |\alpha|^2 = 1$. The notation $|l, m, n\rangle$ describes the state of $l$ UV-photons, $m$ signal-photons and $n$ idler-photons. Each s-i pair exits the crystal in the form of two intersecting cones – figure 2. From these cones, the screens E and E' select by two small holes in each screen, thin fascicles, one with the signal photon and one with the idler photon. The down-conversion effect has a very small probability. Besides that, the intensity of the laser beam is adjusted so as to have only one signal-idler pair in the apparatus at a time, i.e. if one s-i pair is detected at a time *t*, the next pair is detected after a time interval that exceeds the pair coherence time [23].

**Experiment I)** Behind the screens E and E' are placed pairs of detectors, S, I, and S', I', respectively – figure 1. The detectors are considered ideal. We will denote by $\eta^2$ the transmission coefficient of the two screens. Thus, the probabilities of pair detection are

$$P_{\mathrm{S,I}} = P_{\mathrm{S',I'}} = ½\,|\alpha|^2|\eta|^2 . \tag{3}$$

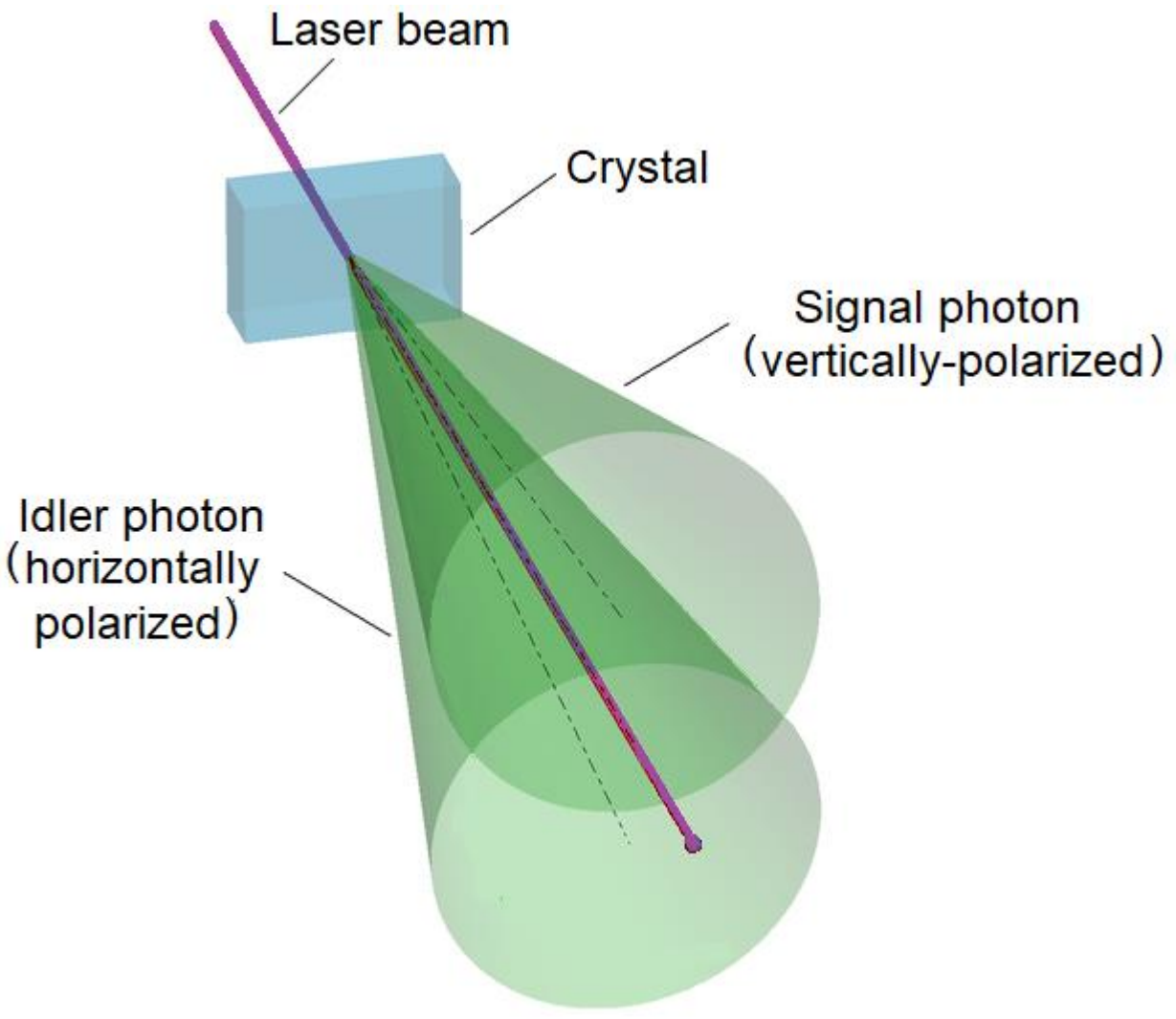

Figure 2. Down-conversion emission.
A typical tableau of down-conversion pairs, the signal and the idler photon belonging to the surface of different cones.

The total probability of detection of a s-i pair is

$$P^0_{\mathrm{detection}} = |\alpha|^2|\eta|^2 . \tag{4}$$

The meaning of the superscript '0' is the s-i pairs emitted by the two crystals are recorded separately.

**Experiment II)** The experiment described below illustrates the effects of two-particle interference. It is similar with the experiment performed by Herzog et al., **[24]**, however the purposes are different. While the experiment in **[24]** had the purpose of showing that the two-particle interference can increase or decrease the number of two-photon pairs, as shown further in section 3, the purpose here is to use this effect for investigating the idea of full and empty waves.

The detectors S', I', from the figure 1 are removed, and the laser is relocated, figure 3. The signal photon and the idler photon from the screen E' are deflected by mirrors M so as to cross one another on the surface of the crystal X, in the place where the beam $|1_p\rangle$ lands on the crystal. The mirrors M are positioned in such a way that the optical path between the two crystals be equal for the beams $|1_{s'}\rangle$ and $|1_{i'}\rangle$. The laser and the mirrors are positioned such that the path length from the beam-splitter BS to the crystal X along the beam $|1_p\rangle$, be the same as along the beam $|1_{p'}\rangle$ and $|1_{s'}\rangle$ (or $|1_{p'}\rangle$ and $|1_{i'}\rangle$). Thus, the s-i wave $|1_{s'},1_{i'}\rangle$ and the beam $|1_{\mathrm{p}}\rangle$ reach the crystal X simultaneously. On the path of $i'$ is placed a phase-shifter by $\phi$. It will be shown below that this phase-shift controls the yield of pairs from the crystal X.

The probability of pair emission from X' and of transmission through the screen E' does not change with the change in trajectory of the photons exiting E' . The state of the system on the input side of X is therefore

$$|\Psi\rangle = \frac{1}{\sqrt{2}}\left\{\beta\,|1_{c'},0,0\rangle + \alpha\,\eta\,\mathrm{e}^{\mathrm{i}[2\pi(l_{p'}+l_{s'}+l_{i'})/\lambda+\phi]}\,|0,1_{s'},1_{i'}\rangle + \mathrm{e}^{\mathrm{i}2\pi l_p/\lambda]}\,|1_p,0_s,0_i\rangle\right\}, \tag{5}$$

where $l_{p'}$, $l_p$, $l_{s'}$, and $l_{i'}$, are the path-lengths of the beam $|1_{p'}\rangle$ from BS to X', of the beam $|1_p\rangle$ from BS to X, of the signal $|1_{s'}\rangle$ and of the idler $|1_{i'}\rangle$ from X' to X. We will denote

$$l_{p'} + l_{s'} + l_{i'} = L' . \tag{6}$$

The UV photon $|1_p\rangle$ produces in the crystal X an s-i pair in the form of intersecting cones – figure 2 and transformation (1) – so, the w-f (5) becomes

$$|\Psi\rangle = \frac{1}{\sqrt{2}}\left\{\beta\,|1_{c'},0,0\rangle + \alpha\,\eta\,\mathrm{e}^{\mathrm{i}(2\pi L'/\lambda+\phi)}\,|0,1_{s'},1_{i'}\rangle + \mathrm{e}^{\mathrm{i}2\pi l_p/\lambda}\big(\beta\,|1_c,0_s,0_i\rangle + \alpha\,|0_c,1_s,1_i\rangle\big)\right\}. \tag{7}$$

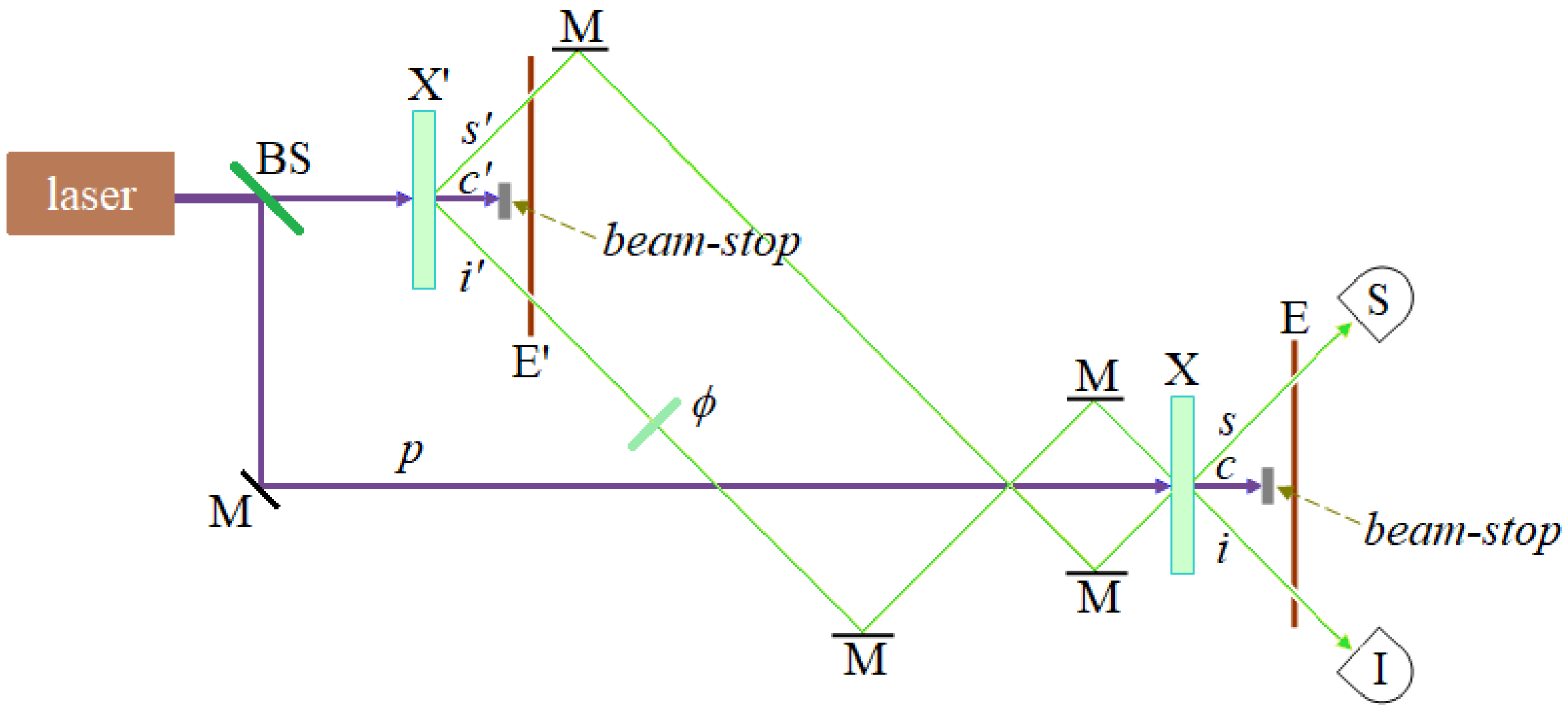

**Figure 3.** Bringing the paths of the DC-photons from two crystals to overlap.
See explanations in the text.

From these cones the screen E cuts two fascicles, so the w-f (7) evolves into

$$|\Psi_1\rangle = \frac{1}{\sqrt{2}}\{\beta(e^{i2\pi l_{p'}/\lambda}|1_{c'},0,0\rangle + e^{i2\pi l_p/\lambda}|1_c,0_s,0_i\rangle) + \alpha\,\eta[e^{i(2\pi L'/\lambda+\phi)}|0,1_{s'},1_{i'}\rangle + e^{i2\pi l_p/\lambda}|0_c,1_s,1_i\rangle]\}. \tag{8}$$

The 2-wave $|1_{s'},1_{i'}\rangle$ incident to X is partially up-converted in X to UV photons. The rest of this wave passes the crystal unperturbed and is truncated by the screen X which cuts off the tails of the two fascicles. Let ξ be the up-conversion amplitude and $1/\gamma$ the reduction in amplitude of this 2-wave by the screen E, i.e.

$$|0,1_{s'},1_{i'}\rangle \xrightarrow[X]{} \xi\,|1_c,0_s,0_i\rangle + \gamma\sqrt{1-|\xi|^2}\,|0_c,1_s,1_i\rangle\,. \tag{9}$$

Introducing this transformation in (8) and arranging terms, one gets

$$|\Psi_2\rangle = \frac{1}{\sqrt{2}}\{\alpha\eta[\gamma\sqrt{1-|\xi|^2}\,e^{i(2\pi L'/\lambda+\phi)} + e^{i2\pi l_p/\lambda}]|0_c,1_s,1_i\rangle + \beta\,|1_{c'},0,0\rangle + [\alpha\,\eta\xi\,e^{i(2\pi L'/\lambda+\phi)} + e^{i2\pi l_p/\lambda}\beta]|1_c,0_s,0_i\rangle\}. \tag{10}$$

The upper line on the RHS expresses what impinges on the detectors S and I,

$$|\Phi\rangle = \frac{1}{\sqrt{2}}\alpha\eta[\gamma\sqrt{1-\xi^2}\,e^{i(2\pi L'/\lambda+\phi)} + e^{i2\pi l_p/\lambda}]|0_c,1_s,1_i\rangle\,. \tag{11}$$

## 3. The enhanced and the inhibited emission of pairs

In writing the expression on the RHS of (11) the fact was taken into account that beyond the screen E, nothing reminds the origin of the pair, i.e. whether it was born in the crystal X', or in the crystal X. In consequence, two-particle interference occurs between these two two-particle waves.
Let's make the coarse approximation,

$$\xi = 0, \gamma = 1, \tag{12}$$

see the transformation (9), i.e. the wave $|0, 1_{s'}, 1_{i'}\rangle$ passes through the screen E as is. Then, (11) becomes

$$|\Phi\rangle = \frac{1}{\sqrt{2}} \alpha \eta e^{i2\pi l_p / \lambda} \left[ e^{i[2\pi (L' - l_p)/\lambda + \phi]} + 1 \right] |0_c, 1_s, 1_i\rangle . \tag{13}$$

Two cases are particularly interesting:

**A)**

$$\phi = -2\pi (L' - l_p) / \lambda . \tag{14}$$

From the RHS of (13) results

$$P^{+}_{\text{detection}} = 2|\alpha|^2 |\eta|^2 . \tag{15}$$

The meaning of the upper script '+', is *enhanced yield of pairs*, twice more than in the case when the pairs $|1_{s'}, 1_{i'}\rangle$ and the pairs $|1_s, 1_i\rangle$ were detected separately. Since, as said above, the yield from the crystal X' is the same as in the experiment I, it means that the crystal X produces more pairs than in the experiment I.

**B)**

$$\phi = \pi - 2\pi (L' - l_p) / \lambda . \tag{16}$$

From the RHS of (13) there results

$$P^{-}_{\text{detection}} = 0 . \tag{17}$$

The meaning of the upper script '–' is *inhibition of pair-production*. As one can see at an examination of the w-f in (8) and (9), the fascicles of photons coming from the screen E' interfere destructively with the cones from E within these fascicles.

## 4. The full/empty waves hypothesis

In this section is examined the meaning of the two cases presented above, under the assumption of full and empty waves. These two types of wave can be added or subtracted from one another, can produce interference, can participate in any experiment, however, when brought to detectors only the full waves make them click.

We start with the case **A**.
The production of pairs by the crystal X' and the transmission probability by the screen E' do not change with the change in the configuration from the figure 1 to the figure 3. Therefore, with the approximations (12), the probability that a pair originating in X' reach and impress the detectors S and I, remains $P_{S',I'} = ½ P^0_{detection}$. As said in the introduction of the section 2, the apparatus is tuned so that $P^0_{detection}$ is equal to no more than one s-i pair detected in a single trial of the experiment. For simplicity, let's discard the trials ending without detection. Thus, in two consecutive trials, on average, one of the pairs detected by S and I was born in the crystal X'.
In the experiment I was detected another pair, once in two trials, produced in X without the assistance of the pair from X'. However, as shown by (15), the detection probability $P^+_{detection}$ is twice greater than $P^0_{detection}$, i.e., on average, two pairs are detected per trial. Let's examine the process of generation of the extra-pairs.

After an s-i pair is detected by the detectors S and I, it leaves the crystal X. So, if the pair came from X', after reaching the detectors it is no more present in the crystal.
The additional pair per trial is conditioned by the phase condition (14) which needs the presence of the 2-wave from X'. According to the QM the time at which an s-i pair is born in X' is not known. However, whichever would be this time, in one of two trials on average there is no pair coming from X' to X. So, on average, the condition (14) should be satisfied in one of two trials. The yield of pairs should be three pairs per two trials, in one trial a pair should come from X' and another pair should be generated in X due to the condition (14), and in another trial a pair should be born in X without the assistance of the pair from X'.
But this is not what happens. It seems that the pair from X' is present in each trial. That may be explained in terms of empty waves. In one of two trials the pair from X' comes both as a full 2-wave and produces a detection, and is also present for fulfilling the condition (14) for the generation of the extra-pair. In the other trial the pair from X' comes only as an empty wave.

In the case **B**,
the 2-waves, one from E' and one generated in X within the fascicle coming from E', are destroyed in the crystal X according with the phase condition (16). However, as shown in the case **A**, the s-i pair from X' comes only in one of two trials. Thus, the condition (16) can be fulfilled in one of two trials. Therefore, in the other trial a pair should be born in X and detected by the detectors S and I.
But this is not what happens. Here too, the explanation can be given by means of empty waves. The 2-wave from X' comes in one trial as a full wave, and in the other trial as an empty wave. The full wave is up-converted because it meets an empty 2-wave produced in X. The empty 2-waves from X' and from X annihilate one another.

## 5. Conclusion

An experiment was described and analyzed in general according to the quantum formalism, and, also under the assumption of full and empty waves. Two cases were given special attention, one in which the yield of s-I pairs is inhibited, and one in which the yield is enhanced. It was shown that for these cases, the hypothesis of full and empty waves seems necessary and provides a very plausible explanation.